\documentclass{nature}
\usepackage{ae}
\usepackage[pdftex]{graphicx}
\usepackage{amsmath}
\usepackage{amssymb}
\usepackage{epsfig, float}
\usepackage{hyperref}

\title{Generalized Gelation Theory describes Human Online \\ Aggregation in support of Extremism}
\author{Pedro D. Manrique, Minzhang Zheng, Zhenfeng Cao \& Neil F. Johnson}

\begin{document}
\maketitle

\begin{affiliations}
\item Physics Department, University of Miami, Coral Gables, Florida FL 33126, U.S.A.
\end{affiliations}

\begin{abstract}
Though many aggregation theories exist for physical, chemical and biological systems, they do not account for the significant heterogeneity found, for example, in populations of living objects \cite{Donati06,Bounova12,Klett78,Wall80,Wattis96,Brock70,Stockmayer43,Benz01}. This is unfortunate since understanding how heterogeneous individuals come together in support of an extremist cause, for example, represents an urgent societal problem.  
Here we develop such a theory and show that the intrinsic population heterogeneity can significantly delay the gel transition point and change the gel's growth rate. We apply our theory to examine how humans aggregate online in support of a particular extremist cause. We show that the theory provides an accurate description of the online extremist support for ISIS (so-called Islamic State) which started in late 2014. \end{abstract}


Physical theories of aggregation rely on kinetic equations where two clusters of sizes $i$ and $j$ join to form a new cluster of size $i+j$ at a rate $K_{ij}$, also referred to as the kernel \cite{Stockmayer43,Brock70,Flory53}. The conventional theory proposes a set of rate equations for the number of clusters of size $s$, $n_s(t)$, for $s=1,2,...$. This deterministic analysis is known as Smoluchowski theory \cite{Wattis06,SMT1,SMT2}. It is known that if the aggregation rate increases rapidly with the cluster sizes, the system undergoes a large-scale transition where a non-negligible fraction of the total population aggregates into the largest cluster, known as a {\em gel} \cite{RednerBook,Ziff82,Lushnikov06}. A kinetic model of polymerization allows for analytical treatment for a few types of kernels. In particular, it has been shown that a kernel of the form $K_{ij}\sim (ij)^{\omega}$ yields a gel transition for $\omega>1/2$, and the system's size distribution at the point of transition follows a power-law (PL) with exponent $\tau=3/2+\omega$ \cite{Ernst82,Ernst83}. Such aggregates may subsequently fragment, however our focus in this paper is on how they initially emerge and grow, and the consequences of this. 
The theory that we now develop shows that heterogeneous systems with aggregation rules based on objects' mutual affinity, can effectively delay the gel transition point and drastically alter its growth rate. We then apply our theory to analyze the formation and early dynamics of a collective human support network underlying extremism. Our study provides new insights into the potential mechanisms underlying their online aggregation dynamics. More generally, our theory can be applied to an aggregating population of any type in which there is significant individual heterogeneity.
\begin{figure}
\centering
\includegraphics[width=0.95\linewidth]{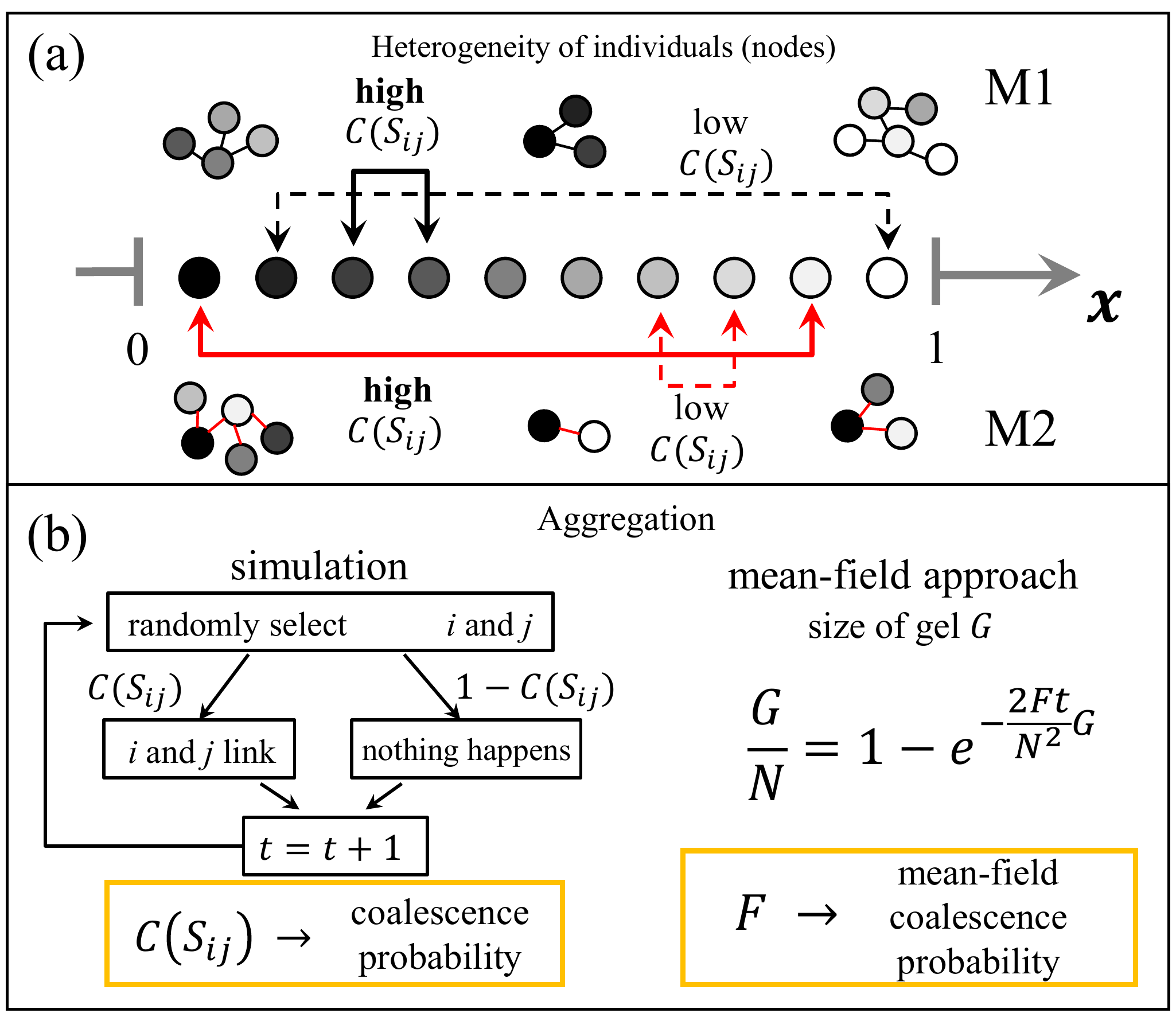}
\caption{\small{{\bf Our model of aggregation in a heterogeneous population} (a) Individual heterogeneity is modeled by introducing a unique hidden variable $x$ to each agent. The value of $x$ is randomly assigned from a given distribution $q(x)$. The formation mechanism depends on the affinity among individuals. Favoring high similarity in establishing a connection is mechanism 1 (which we refer to as M1) while favoring dissimilarity is mechanism 2 (which we refer to as M2). These mechanisms are quantified by the coalescence probability $\mathcal{C}(S_{ij})$, which is a function of the similarity $S_{ij}=1-|x_i-x_j|$ of a given pair of objects $i$ and $j$. (b) The aggregation process and hence the formation of the gel cluster ($G$) can be described by a discrete simulation (left panel) and analytically traced by a mean-field approach (right panel). }}
\end{figure}

We incorporate the heterogeneity among objects by a single variable $x$ that for simplicity we refer to as its `character' and which is  assigned to each individual \cite{Char1,Char2}. For simplicity, we consider $x$ to be a real number between 0 and 1 given that any other one-dimensional range can be easily rescaled to this. In addition, we assume that $x$ is static over time, though this aspect could be changed in the future to account for experience or external influence. For a given population of $N$ objects, a distribution of character values $q(x)$ is used to randomly assign a unique $x_i$ to each object $i$, ($i=1,2,3,...,N$). Interactions between objects are described by means of their mutual affinity. We define the similarity $S_{ij}$ between individual $i$ and individual $j$ as $S_{ij}=1-|x_i-x_j|$, so that individuals with alike character have a high similarity while individuals with unlike character have a low similarity. Hence we illustrate two contrasting aggregation mechanisms: Mechanism 1 (which we refer to as M1) forming alike clusters and Mechanism 2 (which we refer to as M2) forming unlike clusters. Figure 1(a) schematically illustrates these aggregation mechanisms that are quantified by the coalescence probability $\mathcal{C}(S_{ij})$ between individuals $i$ and $j$.

Starting from an isolated population of $N$ objects, clusters form over time by randomly selecting two individuals $i$ and $j$ that merge into a new cluster with a probability $\mathcal{C}(S_{ij})$ or remain separated with a probability of $1-\mathcal{C}(S_{ij})$. A diagram of the aggregation process is shown in the left panel of Fig 1(b). This process can be traced analytically by means of Smoluchowski theory where two clusters join at a rate proportional to the product of their sizes (i.e. product kernel) and is weighted by the mean-field aggregation probability $F$ that accounts for the heterogeneity and formation mechanism. In general, $F$ determines the likelihood for any pair of elements $i$ and $j$ to merge into a new cluster at a given timestep $t$. The set of coupled differential equations for the dynamics of the number of clusters of size $s$, $n_{s}(t)$, is given by:
\begin{eqnarray}
\dot{n}_{s}(t)&=&-2F\frac{sn_{s}}{N^2}\sum_{r=1}^{\infty}{rn_{r}}+\frac{F}{N^2}\sum_{r=1}^{s}rn_{r}(s-r)n_{s-r}, \ \ \quad s\geq 2\\
\dot{n}_{s}(t)&=&-2F\frac{n_{s}}{N^{2}}\sum_{r=1}^{\infty}rn_{r},\quad s=1\ \ .
\end{eqnarray}
The first term on the right-hand side of both Eqs. 1 and 2 represents the population of clusters of size $s$ that merge with other clusters, while the second term in Eq. 1 is the population of smaller clusters that merge to form clusters of size $s$. Since $K_{ij}$ is a product kernel with exponent $\omega=1$, the system undergoes a gelation transition at some finite point in the dynamics \cite{Ziff82}, and follows a PL size distribution with exponent $\tau=5/2$ at the gel transition point $t_c$. Its exact location is determined mathematically by a singularity in the second moment of the size distribution and found to be equal to $t_c=N/2F$ (see Supplementary Information for the full derivation). 

The expression for the evolution of the gel cluster $G$ can be obtained by means of the exponential generating function $\mathcal{E}(y,t)\equiv\sum_{s\geq1}sn_{s}e^{ys}$ whose partial time derivative takes the form of the inviscid Burgers equation which can be solved by the method of characteristics (see Ref. \cite{RednerBook} for the case of homogeneous systems). Above the gel transition point, the formalism determines that the size of $G$ follows the following equation
\begin{equation}
G(t)=N\left(1-e^{-\frac{2Ft}{N^2}G(t)}\right).
\end{equation}
The solution of (3) can be expressed in terms of the $W$-Lambert function as $G/N=1-W(z\exp z)/z$ where $z=-2Ft/N$.

The aggregation mechanism favoring joining similar individuals (i.e. M1) is defined simply as $\mathcal{C}=S_{ij}$, while for aggregating dissimilar individuals (i.e. M2) it is defined as $\mathcal{C}=1-S_{ij}$. The limit of random aggregation is obtained by considering all character values to be identical (i.e. $q(x)=\delta(x-x_{0})$) which makes the process character-free with an aggregation probability $\mathcal{C}(S_{ij})=1$. For the case of interest of a heterogeneous population, the mean-field probability $F$ depends on both the formation mechanisms and the character distribution. For example  for a uniform character distribution $q(x)$, the probability density function (PDF) of the similarity $y=S_{ij}$ for M1 is $f(y)=2y$ and hence the mean-field aggregation probability $F=\int_{0}^{1}yf(y)dy=2/3$. Similarly, for M2 the mean-field probability results in $F=1/3$. The homogeneous (i.e., character free) limit occurs when $y=1$ and the character distribution is a Dirac delta which yields $F=1$. 

Figure 2(a) summarizes these parameters as well as illustrating a single simulation result of the gel formation (before and after the theoretical transition point $t_c$) for each of the mechanisms. The disks represent the evolution of $G$ while the rings are smaller clusters whose radii are proportional to the square root of their respective size. Figure 2(b) compares the time evolution of $G$ for each of the aggregation mechanisms calculated from the mean-field approach (solid lines) and the discrete simulations (dots) averaged over 500 realizations. They are in good agreement.

\begin{figure}
\centering
\includegraphics[width=0.99\linewidth]{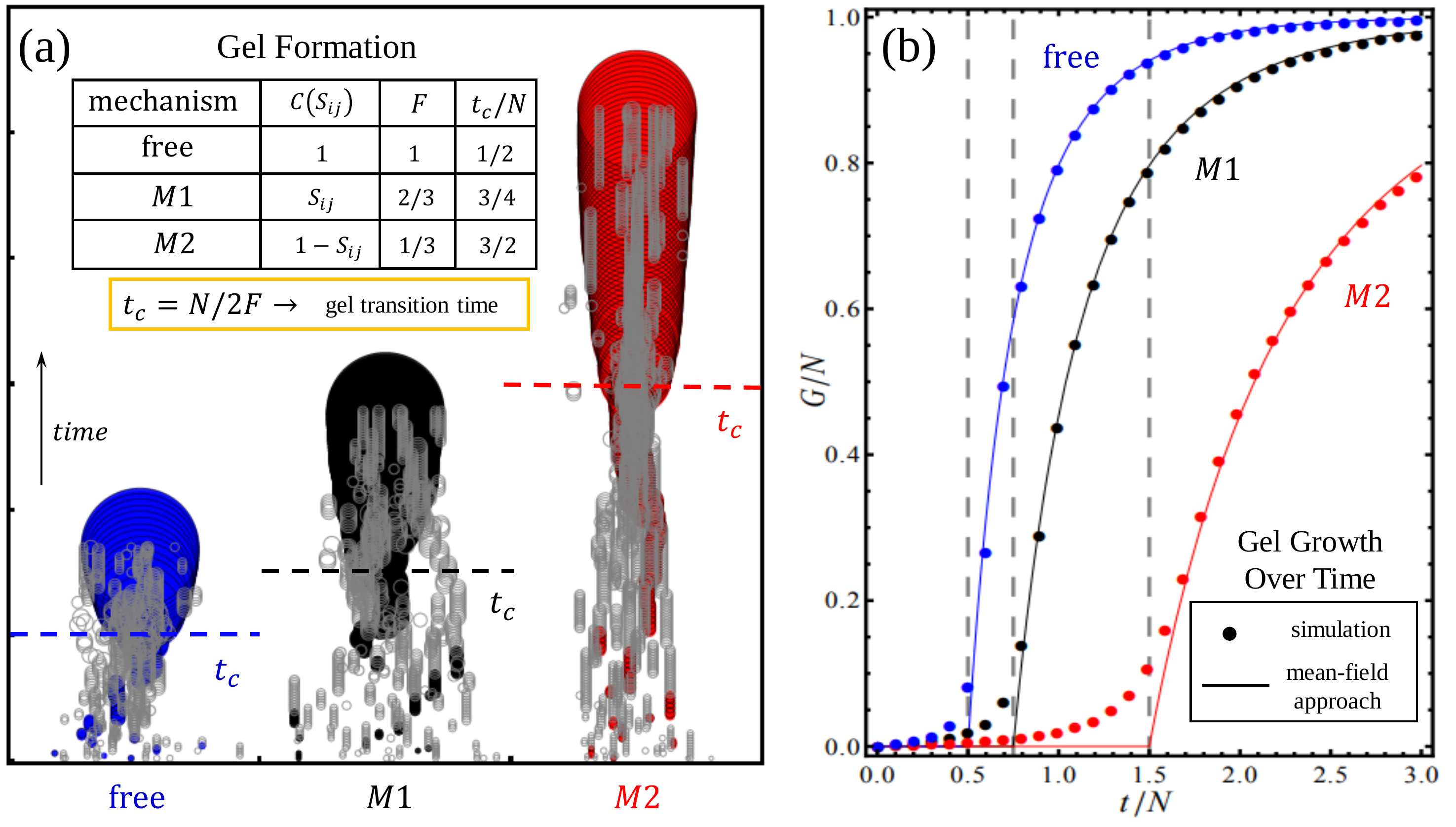}
\caption{{\bf Stochastic and mean-field approach results for the gel dynamics} (a) Formation rules and parameters for each mechanism (free, M1 and M2) and a sample simulation of the dynamics. The 'free' mechanism refers to a homogeneous system with all $x$ values identical. Colored disks represent the evolution of $G$ while the gray ones are smaller clusters. For all cases the radii grow proportional to $s^{1/2}$ and the time limit is set when $G$ reaches $70$\% of $N$ ($N=10^3$ agents). Dashed horizontal lines indicate the theoretical transition time $t_c$. (b) Contrast between simulation (points) and mean-field approach (lines) for the evolution of $G$ for the proposed mechanisms.}
\end{figure}

Our model can be used to analyze the evolution of heterogeneous systems in the natural world as well as social domains. As an illustration, here we study the formation of online extreme clusters (henceforth referred to as `groups' since they are online social media groups) each of which comprises a finite number of individual human users. 
Such online group formation in support of Islamic State (ISIS) through VKontakte (VK, http://vk.com) has been reported in previous studies \cite{VKScience16,VKOrg17}. VKontakte is the largest European social media platform and is based in Russia. As of September of $2017$, VKontakte had $447$ million users worldwide and is known to have been heavily used to spread pro-ISIS propaganda as well as ISIS recruitment and financing \cite{VKScience16,BBC15-1}. Figure 3(a) shows a snapshot of the pro-ISIS network extracted on January 10th 2015: 59 different social media groups supporting ISIS were found, with a total of 21,881 followers combined and 48,605 links (i.e. follows). As a result of the extreme content shared in these groups, moderators are constantly chasing them and shutting them down \cite{BBC15-2,Berger16,Paraszczuk15}. Unlike platforms like Facebook which shuts down these groups almost immediately, VK can takes weeks or even months to act. This allows us to study their rich dynamics. During the period between the end of 2014 and the beginning of 2015, a sudden and roughly continuous growth in the number of added links (i.e. follows) within the whole network occurred and lasted until mid-2015 where a decay process set in \cite{VKOrg17}. The first few weeks of this sudden growth are particularly interesting since the number of shutdown events was minimal and hence aggregation processes dominated the system dynamics. The monitoring of the group size distribution over time revealed that, for three consecutive days starting December 28, 2014, a PL distribution with exponent near $5/2$ cannot be rejected ($p\approx 0.64$). 

\begin{figure}
\centering
\includegraphics[width=0.99\linewidth]{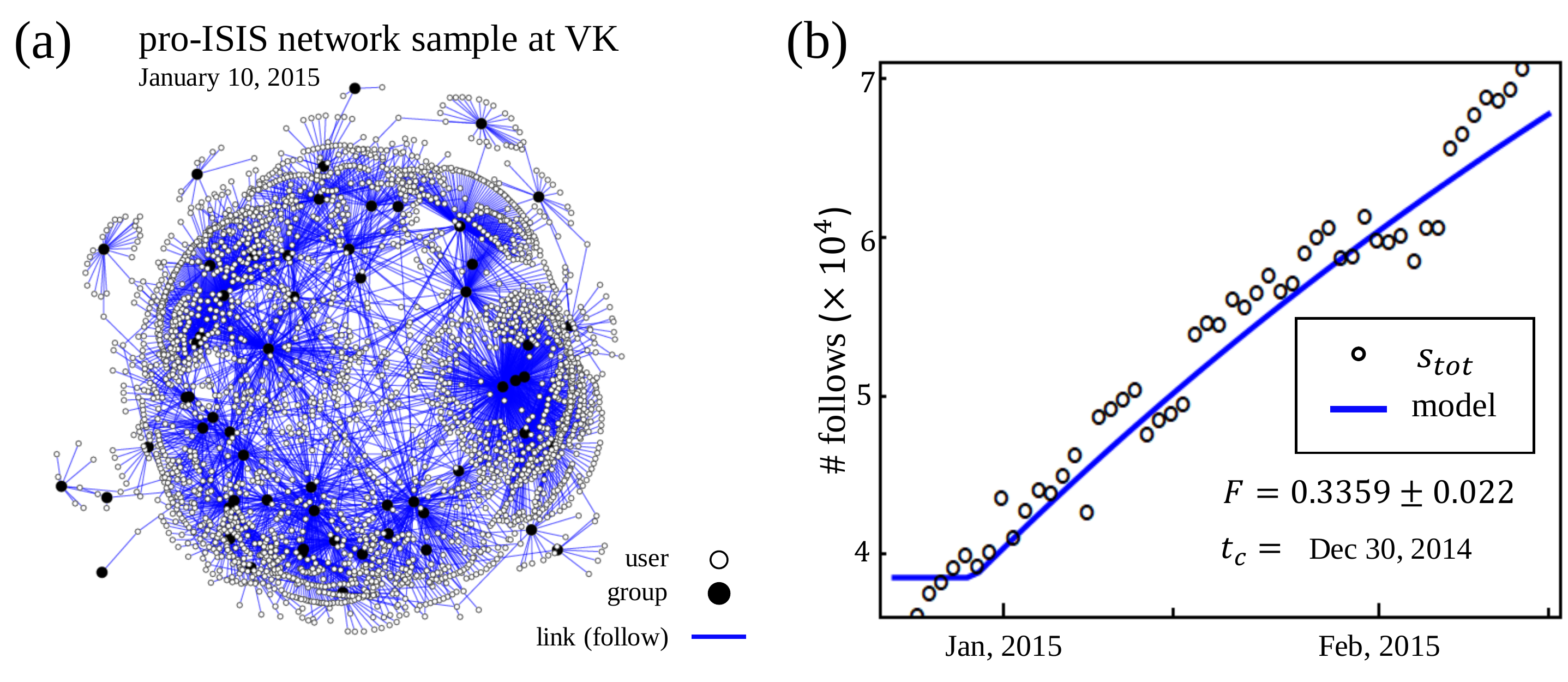}
\caption{\small{{\bf Rise of global pro-ISIS online support} (a) Sample of the online network of groups in support of Islamic State on the VKontakte (VK) platform for an example day: January 10, 2015. Black dots represent groups, white dots are users which are connected through blue links (i.e. follows). (b) Evolution of the total number of follows $s_{tot}$, i.e. links in the bipartite network shown in (a) (black circles), compared to our analytical model of heterogeneous objects undergoing a gelation process (blue line). The transition point $t_c$ is found to be December 30, 2014 which is the time where the size distribution of online groups is approximately $5/2$ ($\alpha=2.46$). A fitting of Eq. 3 yields $F=0.3349\pm0.022$.}}
\end{figure}

These empirical findings suggest that the sudden growth of online pro-ISIS support can be interpreted as a gelation transition within the online network of global users. To gain a deeper insight, we compare the network growth with the gelation curve (Eq. 3) when the transition point corresponds to the third day of the 3-day PL finding ($t_c$=December 30, 2014). Figure 3(b) shows that our heterogeneous model compares well with the real data for a mean-field aggregation probability of $F=0.3359\pm0.022$. The time unit in the model is scaled to that of the data by means of the network's growth rate during a period of $20$ days around the transition point. Since the system was active for years prior to $t_c$, during which multiple aggregation and fragmentation events would have taken place, it is understandable that a certain level of noise is present within the data. We consider this background noise as the floor from which the gel cluster arises at $t=t_c$, as shown in Fig. 3(b). Our results suggest that dissimilar follows collectively assemble to create the network of pro-ISIS support groups in VK (i.e. pro-ISIS support is dominated by mechanism M2). This finding is consistent with the fact that different online social media groups support pro-ISIS causes in complementary ways, such as financing, recruitment, spreading of propaganda among others \cite{VKScience16,BBC15-1}.

Going deeper into the group formation analysis, we now look into the dynamics of individual groups. From the wide ecology of groups found in 2015, we selected those with features that resemble those of our heterogeneous model. In particular, we had to weed out groups that were inactive and/or put their web setting as invisible for long periods of time. In addition, some groups experienced large dynamical increments in very short periods of time which reveals abrupt changes akin to processes such as explosive percolation. Moreover, since the total number of potential follows varies over time, we restrict the modeling to the first few active weeks where the assumption of a constant subpopulation of follows (i.e. $N$) holds. We measured the goodness-of-fit of Eq. 3 with the real data and found a total of 32 groups that give an r-squared higher than $80$\% during the initial growth period. Our approach supposes that each group is a gel cluster formed by a subpopulation of follows from a larger pool comprising the entire network. Our results are presented in Fig. 4. Since groups have the option to turn themselves invisible for any length of time, our data contain some gaps for those particular dates. However we can see that our gelation approach captures well the group growth at the early stage and associates each one with a particular value of $F$ (and hence a formation mechanism) ranging from $1/3$ up to $1$ (see inset of Fig. 4). Specifically, we found that a larger number of groups are formed by similar interests (i.e. M1) than dissimilar (i.e. M2), and for a larger population of groups the mechanism resembles that of a character free model.

\begin{figure}
\centering
\includegraphics[width=0.99\linewidth]{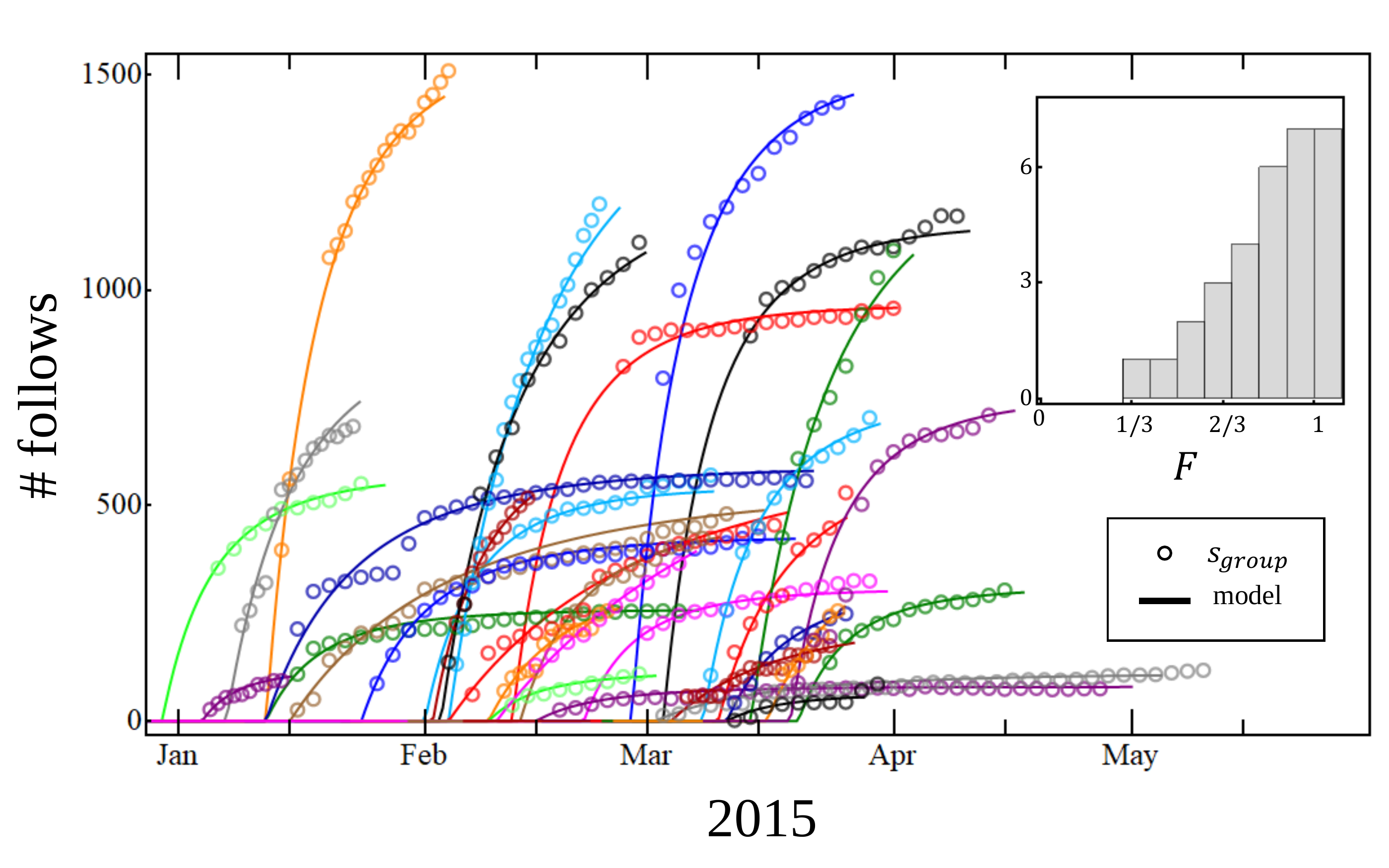}
\caption{\small{{\bf Modeling single online extreme groups}. Evolution of the size of individual online pro-ISIS support groups $s_{group}$ (circles) compared to individual gelation processes (curves) resulting in different values of mean-field coalescence probability $F$ ranging from $1/3$ up to $1$ (inset).}}
\end{figure}

Modeling larger time periods remains a challenge since it involves deriving an expression analogous to Eq. 3 for a population $N$ that changes over time. A first approach to this challenge is by considering small linear variations of $N$ in the original equation. We explore this aspect by adding a small linear increment to the size ($N(t)=N_0+kt$) and comparing to the original case (i.e. $k=0$). Figure S3 illustrates that this variation resembles the group dynamics for a larger time period than the original modeling. Interestingly there are no significant differences in the estimated parameter $F$, which for the static (i.e. $k=0$) and dynamical (i.e. $k\neq0$) versions of the model are $F=0.97\pm0.022$ and $F=0.96\pm0.043$ respectively.

In summary, we have shown that aggregation mechanisms based on individual heterogeneity enrich the dynamics of a finite set of interacting objects while still remaining analytically tractable. We used our theoretical approach to analyze the dynamics of extreme online groups on the global scale as well as the individual group scale. For the overall online dynamics, the sudden rise in the number of follows and size distribution in late 2014 serve as indicators that the system can be interpreted as experiencing a gel transition where a non-negligible fraction of the entire population comprises the extremist network. Our theory goes further by indicating that the underlying aggregation mechanism that better describes the system tends to favor the formation of team-like clusters of dissimilar individuals. This finding agrees with the real data where groups fulfill different but complementary roles in the online network at large. We then looked into the individual group dynamics and found that some groups are better described by means of aggregating similar objects, while others resemble a team-like structure of dissimilar individuals -- and the remainder tend to follow a more homogeneous aggregation process.

\begin{methods}

The full mathematical derivation of Eq. 3 is given in the SI. The power-law analysis that we use for the distribution of group sizes follows strict testing procedures described in the references of the SI. The data extraction is explained in full detail in Ref. \cite{VKScience16} of the main paper.

\end{methods}

\begin{addendum}
 \item[Competing Interests] The authors declare that they have no
competing financial interests.
 \item[Correspondence] Correspondence and requests for materials
should be addressed to P.D.M.~(email: p.manriquecharry@umiami.edu).
 \item[Acknowledgments] We are grateful to Yulia Vorobyeva for detailed help and translation concerning ISIS activity on Russian-language Vkontakte. The authors acknowledge funding from the National Science Foundation grant CNS1522693 and Air Force (AFOSR) grant 16RT0367.
\item[Author contributions] P.D.M. and N.F.J. worked on the mathematical model. P.D.M., M.Z. and Z.C. worked on the data and data analysis. P.D.M and N.F.J wrote the paper. All authors revised and approved the manuscript.
\end{addendum}

\end{document}